\documentclass[sigconf]{acmart}

\AtBeginDocument{%
  }

\copyrightyear{2024}
\acmYear{2024}
\setcopyright{rightsretained}
\acmConference[SA Art Papers '24]{SIGGRAPH Asia 2024 Art Papers}{December
3--6, 2024}{Tokyo, Japan}
\acmBooktitle{SIGGRAPH Asia 2024 Art Papers (SA Art Papers '24), December
3--6, 2024, Tokyo, Japan}
\acmDOI{10.1145/3680530.3695448}
\acmISBN{979-8-4007-1133-6/24/12}

\usepackage[T1]{fontenc}
\usepackage{graphicx}
\usepackage{tabularx} 
\usepackage{booktabs}
\usepackage{ragged2e}
\usepackage{algorithm}
\usepackage{algorithmic}
\usepackage{amsmath}
\usepackage{hyperref}
\usepackage[rightcaption]{sidecap}

\begin{document}

\title[ScribGen]{ScribGen: Generating Scribble Art Through Metaheuristics}

\author{Soumyaratna Debnath}
\affiliation{%
  \institution{CVIG Lab, IIT Gandhinagar}
  \city{Gandhinagar}
  \country{India}
}
\orcid{0000-0002-3690-7216}
\email{debnathsoumyaratna@iitgn.ac.in}

\author{Ashish Tiwari}
\affiliation{%
  \institution{CVIG Lab, IIT Gandhinagar}
  \city{Gandhinagar}
  \country{India}
}
\orcid{0000-0002-4462-6086}
\email{ashish.tiwari@iitgn.ac.in}

\author{Shanmuganathan Raman}
\affiliation{%
  \institution{CVIG Lab, IIT Gandhinagar}
  \city{Gandhinagar}
  \country{India}
}
\email{shanmuga@iitgn.ac.in}
\orcid{0000-0003-2718-7891}

\renewcommand{\shortauthors}{Debnath et al.}

\begin{abstract}
Scribble art, arising from chaos and randomness, remains one of the exceptionally attractive forms of art. Many works bridge the gap between sketches and images, but few translate images into meaningful chaotic expressions. While deep generative networks are known for understanding images, their ability to induce scribble drawings is under-explored. Unlike GAN-based approaches that generate line drawings, sketches, and contours, our work uses metaheuristics to produce scribble art from images. We extensively analyse various metaheuristic algorithms, demonstrating their optimal balance between creativity and computational efficiency. They offer better adaptability and accuracy than state-of-the-art deep generative models for image-to-scribble generation.
\end{abstract}



\begin{teaserfigure}
  \includegraphics[width=\textwidth]{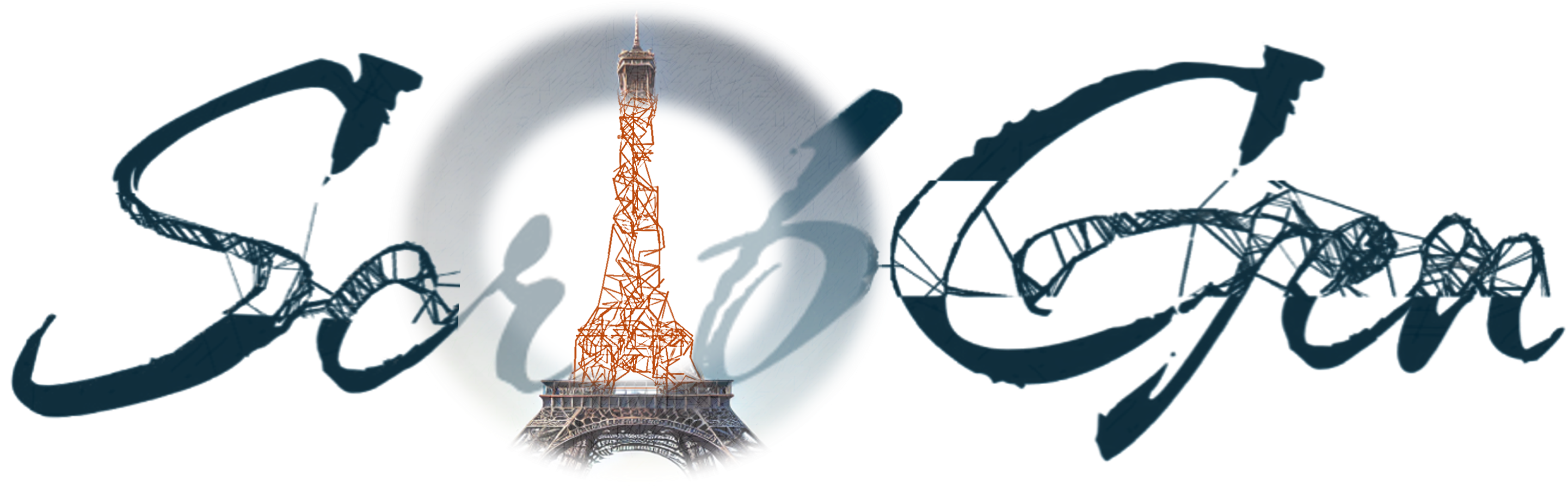}
  \caption{Artistic visual constructed using ScribGen. Input image generated by DALL-E 3 \cite{betker2023improving} from text prompts.}
  \Description{Artistic visuals constructed using ScribGen.}
  \label{fig:teaser}
\end{teaserfigure}

\maketitle

\section{Introduction and Background}

\subsection{Scribble Art}
Art has long been a medium for individuals to engage with the world. Scribble art, a form of abstract visual expression, features spontaneous, gestural strokes made with pens or brushes. These dynamic and expressive compositions, created quickly and impulsively, reveal intricate patterns and hidden meanings upon closer inspection. While scribble art is often associated with spontaneous expression and experimentation, it can also be planned and intentional. Some artists use scribble techniques as a starting point for their creative process, exploring the possibilities of line, shape, and texture before refining their work into more polished compositions. From ancient cave paintings to modern abstract sketches and doodles, scribble art has evolved with civilizations, reflecting diverse artistic movements and cultural influences. This evolution highlights its universal appeal, transcending language and cultural barriers and connecting people through the shared experience of creating art.

\begin{figure*}
\centering
\Description{The figure lists the fundamental ideas of various metaheuristic algorithm categories based on evolution, nature, physics, behaviour and hybrid methods.}
\includegraphics[width=1\textwidth]{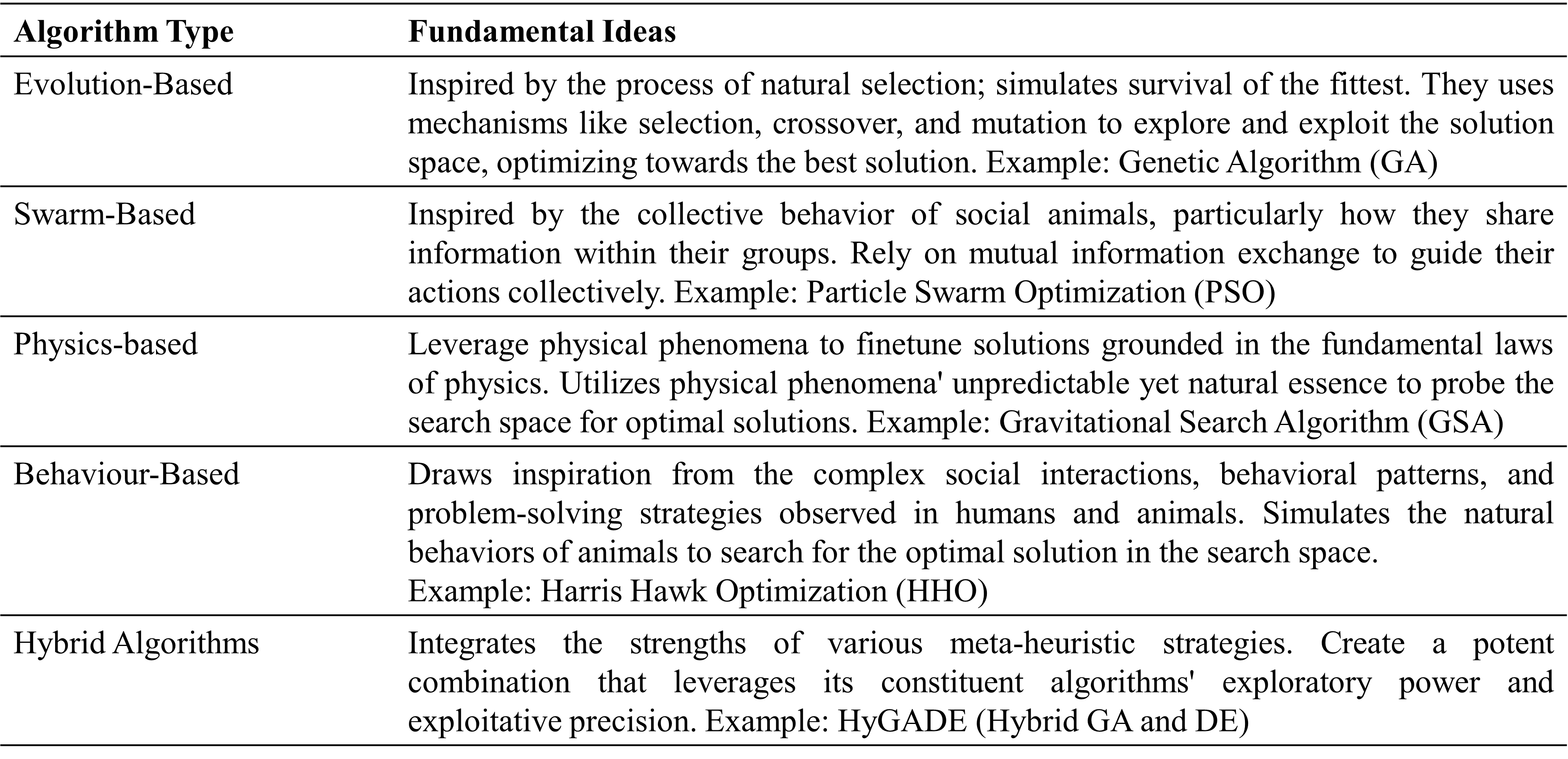}
\caption{Categories of Metaheuristic algorithms} \label{fig:algo_forms}
\end{figure*}

\subsection{Art and Technology} 

Although digital platforms have overshadowed many analogue methods, they have also created new avenues for artistic experimentation. The intersection of art and technology has been explored through deep neural networks, which recreate various art forms, including shadow art \cite{sadekar2022shadow} \cite{gangopadhyay2023search}. Generative Adversarial Networks (GANs) \cite{goodfellow2014generative} have shown promising results in generating realistic images from simple sketches \cite{chen2018sketchygan} \cite{lu2018image}, allowing for computational creativity. Some studies have synthesized circular scribble art from grayscale images \cite{chiu2015tone}, while others generate line drawings from colour images \cite{chan2022learning}. These works have either been limited to circular scribbles (scribble art is much closer to quick linear strokes and random doodles) or by the availability of edge-map-like training data (often missing the artistic touch), thus highlighting the potential for further exploration in creating "meaningfully random" scribbles that mimic human expressions.

\begin{algorithm}[h]
\caption{Proposed Algorithm: ScribGA}
\label{algo:scribga}
\begin{algorithmic}[]
\STATE \textbf{Input:} Population size, $N$ = 100
\STATE \textbf{Input:} Number of Generations, $G_{max}$
\STATE \textbf{Output:} Set of best solutions, $R$
\STATE \textbf{Begin}
\STATE Generate initial population of $N$ solutions $Y_i$ $(i=1,2,...,N)$
\STATE Set the iteration counter $t = 0$
\STATE Initialize $R$ as an empty set
\WHILE{$t < G_{max}$}
    \STATE Compute fitness of each solution as per Figure \ref{fig:fitness}
    \STATE Include the best solution to the set $R$
    \STATE Select parent pairs based on fitness
    \FOR{each pair of parents}
        \STATE Apply crossover to generate offspring
        \STATE Apply mutation to the offspring
    \ENDFOR
    \STATE Evaluate the fitness of the new offspring
    \STATE Replace the old population with the new offspring
    \STATE Increment the iteration counter $t = t + 1$
\ENDWHILE
\STATE \textbf{Return} $R$
\STATE \textbf{End}
\end{algorithmic}
\end{algorithm}

\begin{figure*}[h]
\centering
\Description{The figure represents the process of computing solution fitness and generating scribble art using a generic metaheuristic algorithm backbone.}
\includegraphics[width=0.9\textwidth]{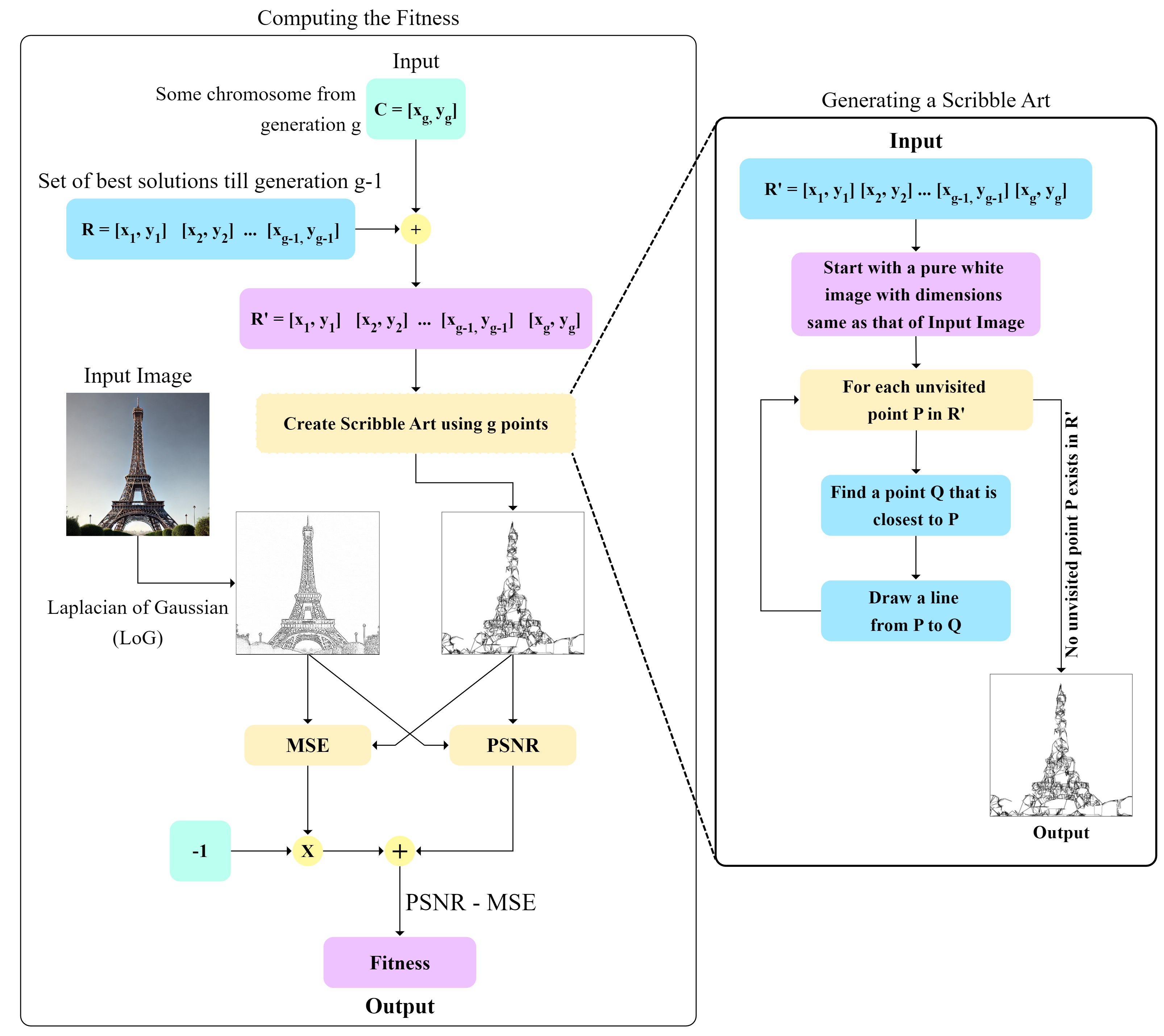}
\caption{Computing solution fitness and generating scribble art. Eiffel tower image generated by DALL-E 3 \cite{betker2023improving} from text prompts.}
\label{fig:fitness}
\end{figure*}

\subsection{Contribution} 

This work explores the potential of using metaheuristic algorithms to generate scribble art. Given a reference image, we define scribble art as an approximation of the image's contours using random yet strategically placed line strokes that preserve the structure of the object in the given image. 

Metaheuristics \cite{osman1997meta} addresses multifaceted, non-linear challenges, thereby identifying high-quality solutions. Unlike classic optimization approaches that rely on gradients, metaheuristics utilize exploration and exploitation-based strategies. These algorithms apply to diverse fields, from mathematics and engineering to healthcare and creative expression. 

In this work, we modify and examine a set of metaheuristic algorithms, i.e., Genetic Algorithm \cite{holland1992genetic} (ScribGA), Differential Evolution \cite{feoktistov2006differential} (ScribDE), Particle Swarm Optimization  \cite{kennedy1995particle} (ScribPSO), Gravitational Search Algorithm  \cite{rashedi2009gsa} (ScribGSA), and Harris Hawk Optimization \cite{heidari2019harris, debnath2023modified} (ScribHHO), to specifically cater to scribble art generation. We also compare our results with learning-based image-to-sketch methods.  

\section{Metaheuristics}

Metaheuristic algorithms initiate with a randomly generated set of candidate solutions, which are iteratively refined based on optimization criteria. They balance exploration and exploitation by initially searching broadly and then focusing on promising areas to find optimal solutions. This process continues until a stopping condition is met, such as reaching a maximum number of iterations or achieving a satisfactory solution.

These algorithms are categorized based on evolution, nature, physics, and behaviour, often working with populations to evolve multiple solutions simultaneously. Figure~\ref{fig:algo_forms} lists the fundamental ideas of various metaheuristic algorithm categories.

\section{Scribble Art Generation as an Optimization Problem}
Generating digital scribble art is an interesting challenge. The goal is to turn a given image into art that looks like free-flowing scribbles while still keeping it recognizable and visually appealing. This task requires balancing randomness and order, with algorithms navigating many possible scribble patterns to find the best representation of the original image.

We propose a novel approach to metaheuristic-based learning tailored specifically for scribble art generation where each solution in the underlying solution space consists of thousands of vertices or edges. We propose a progressive optimization method, where each iteration progressively adds the next best solution from a pool of possible solutions, represented by a 2D point that denotes a position in the image space (pixel). Thus, our objective is to identify a set of points within the image space that, when connected, form a visually compelling yet recognizable scribble drawing.

\begin{figure*}[!htb]
\centering
\Description{The figure illustrates the qualitative effects of different iterations on the Eiffel Tower image using various metaheuristic backbones. Finer details improve progressively with more generations, from 500 to 2000, demonstrating the diversity of these algorithms, much like artists with unique painting techniques.}
\includegraphics[width=\textwidth]{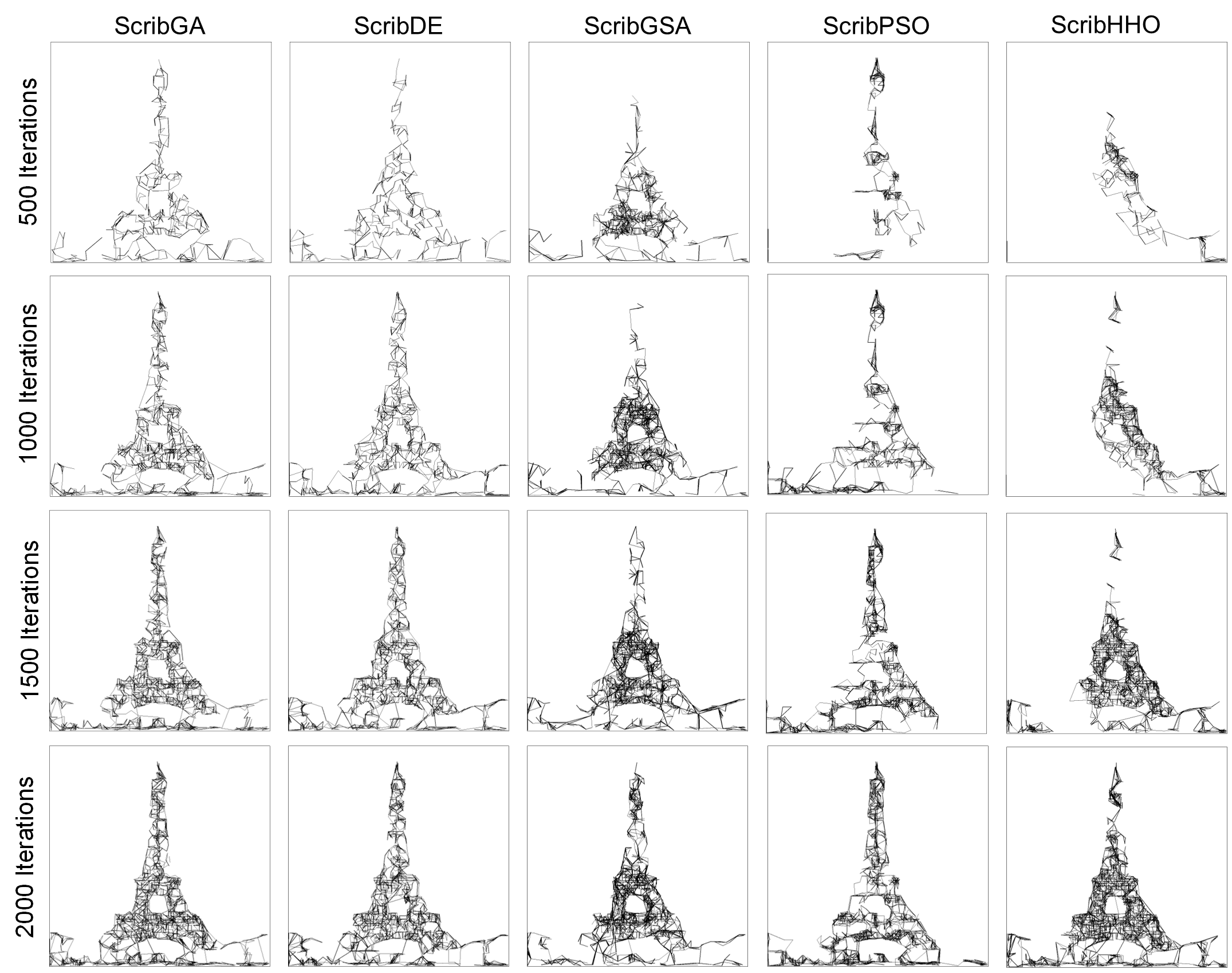}
\caption{Generation-wise scribbling results using different metaheuristic  backbones.} \label{fig:generations}
\end{figure*}

\subsection{Methodology}
Following the framework of a generic metaheuristic backbone algorithm \textit{H}, the process begins with an initial set of solutions \textit{Y}, consisting of \textit{N} randomly initialized points. Each solution is defined by two parameters, \textit{x} and \textit{y}, representing a point within a 2D image-space. We establish a repository \textit{R} to store the optimal solutions identified in each iteration. The selection of these solutions is determined by a fitness score. With each generation, a new optimal point \textit{(x,y)} is added to \textit{R}. After \textit{g} generations, the scribble art generated from the points in \textit{R} represents a refined approximation of the target structure by joining \textit{g} points with \textit{g}-1 line segments. This progressive strategy significantly reduces computational demands while efficiently navigating the complex solution space.

Let \textit{R'} represent the solution state after \textit{g} generations, \textit{I} be the Laplacian of Gaussian (LoG) of the input image, and \textit{I'} be the scribble image formed by the points in \textit{R'}. The idea is to optimize an objective function \textit{F(R')} that maximizes the Peak Signal-to-Noise Ratio (PSNR) and minimizes the Mean Squared Error (MSE) between \textit{I} and \textit{I'} (see Algorithm \ref{algo:scribga} and Figure~\ref{fig:fitness}).

MSE minimizes average pixel differences, while PSNR offers a perceptual perspective aligned with human vision. Using both ensures a balanced optimization, preventing focus solely on error reduction that may not improve visual quality. Together, PSNR and MSE help guide the process towards solutions that combine numerical accuracy with perceptual improvements.

Algorithm \ref{algo:scribga} describes ScribGA - a modified version of the Genetic Algorithm for generating scribbles. Similarly, other variants - ScribDE, ScribPSO, ScribGSA, and ScribHHO, have been reconfigured through a similar approach. 

\begin{figure*}[!htb]
\centering
\Description{The figure presents the qualitative results of scribble drawing generation using various backbone algorithms for the Eiffel Tower, Robot Dog, Taj Mahal, and Portrait images.}
\includegraphics[width=0.95\textwidth]{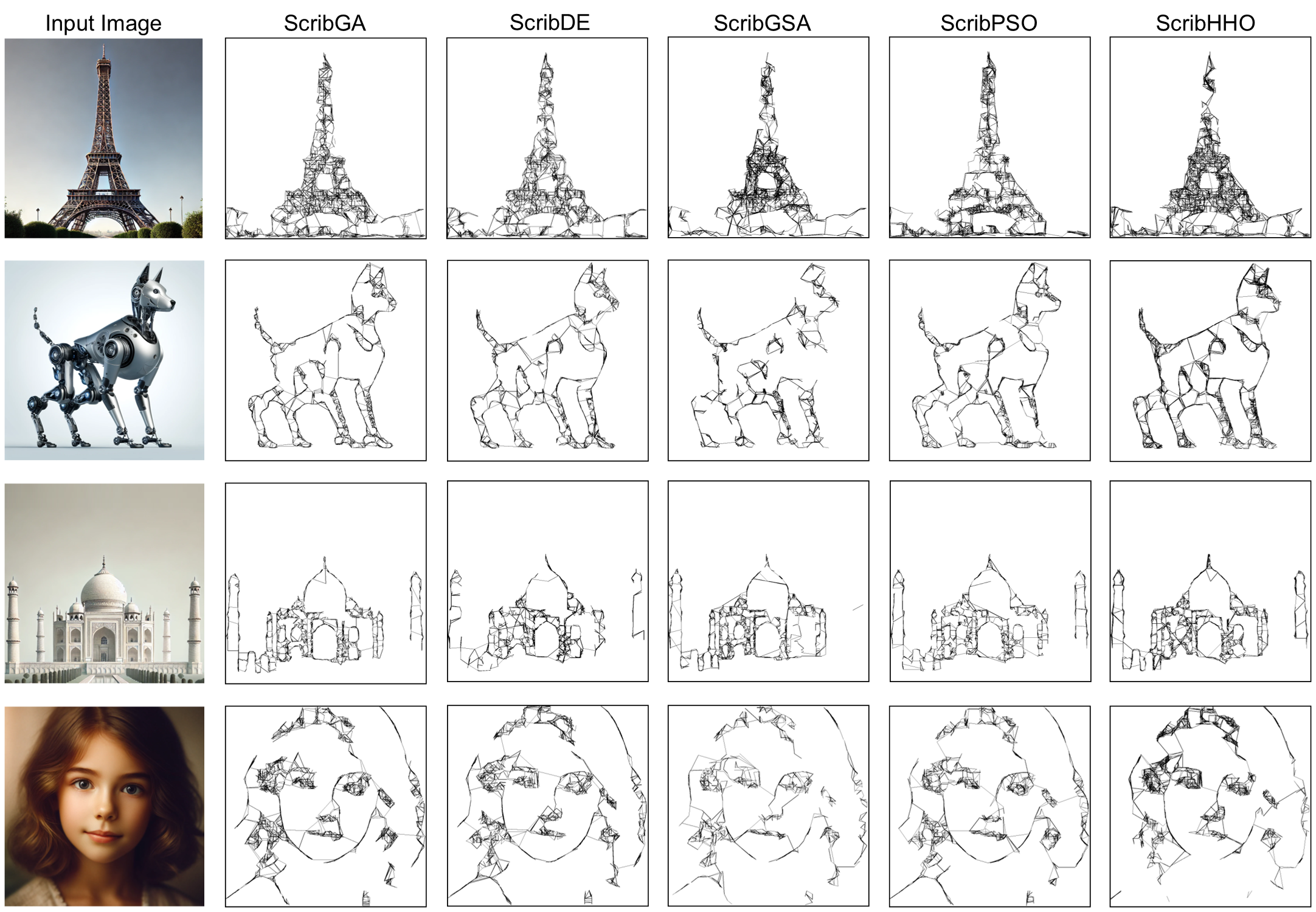}
\caption{Scribbles generated through different metaheuristic backbones after 2000 iterations. Input images generated by DALL-E 3 \cite{betker2023improving} from text prompts.} \label{fig:compared}
\end{figure*}

\begin{figure*}[h]
    \centering
    \Description{The top row of the figure highlights the differences by plotting convergence curves for various test images, showing how each method improves the fitness score over time. The bottom row shows the Structural Similarity (SSIM) between the generated results and the edge maps produced by applying the Laplacian of Gaussian (LoG) to the input images.}
    \includegraphics[width=1\textwidth]{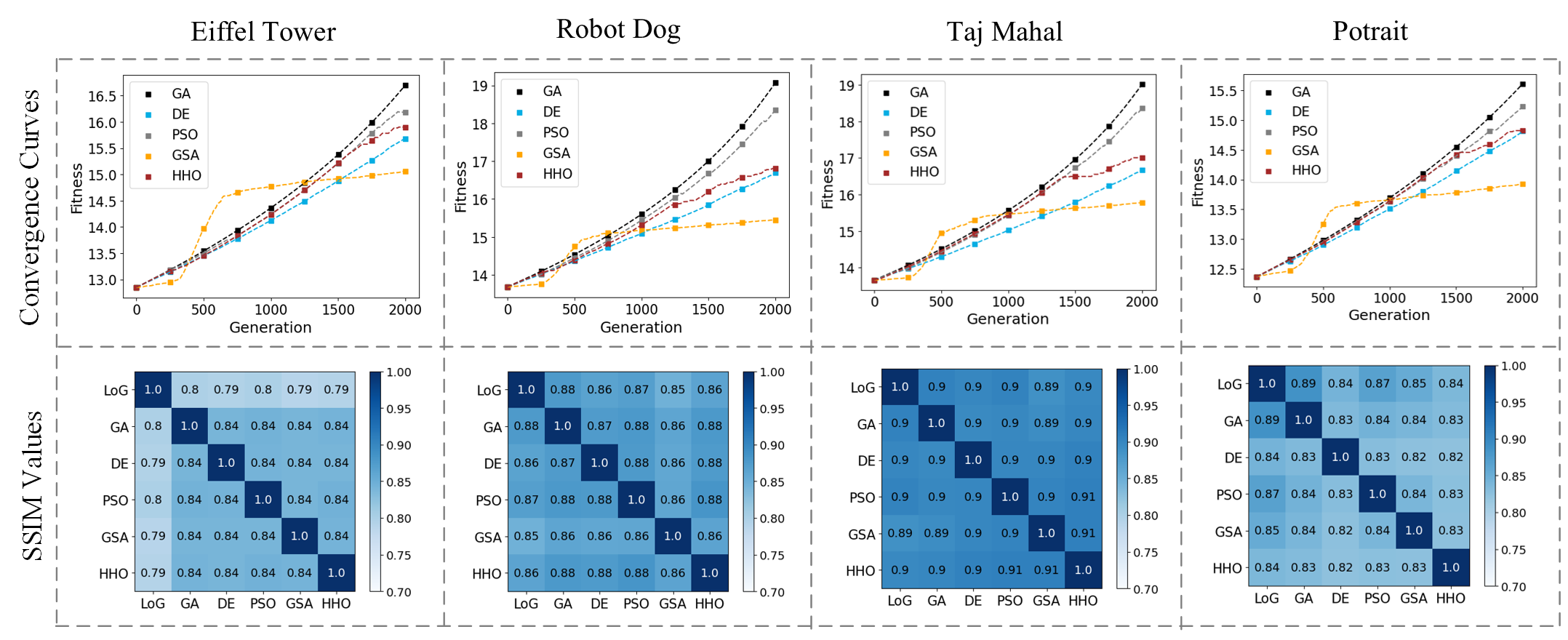}
    \caption{(Top row) Convergence curves and (Bottom row) Structural Similarity index.}
    \label{fig:quant}
\end{figure*}

\begin{figure*}[h]
\centering
\Description{The figure illustrates a comparative analysis of scribbles generated by ScribGen through the continuous evolution over 2000 generations (2000 points) and a composite image derived by combining four distinct images produced in separate runs, each spanning 500 generations.}
\includegraphics[width=1\textwidth]{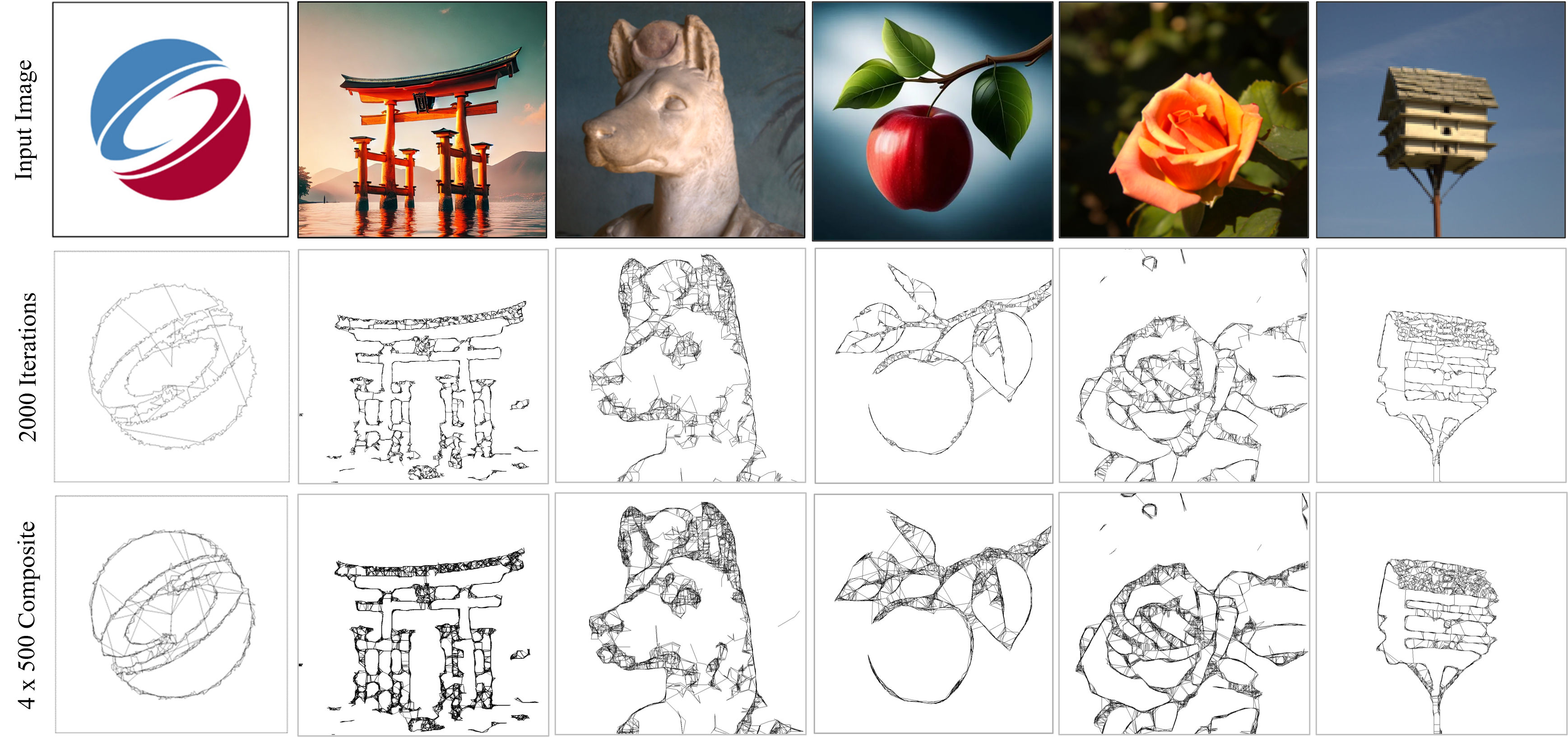}
\caption{Continuous vs composite generative process using ScribGen. Input image credits: SIGGRAPH website [siggraph.org]; COCO dataset \cite{lin2014microsoft}; MIT-Adobe FiveK dataset \cite{fivek}; and DALL-E 3 \cite{betker2023improving} (images generated from text prompts).} \label{fig:combination}
\end{figure*}

\section{Results: Images to Scribbles}

In Figure \ref{fig:generations}, we illustrate the qualitative effects of different numbers of generations (iterations) on the Eiffel Tower image across various metaheuristic backbones. As expected, finer details improve progressively with increasing generations from 500 to 2000. The results show that these algorithms display diverse approaches, just like artists with unique techniques — while some like to work progressively, and others paint in layers. Our supplementary video illustrates how the results evolve across generations.

We observed that ScribGA and ScribDE incrementally refine details across the entire structure, while ScribGSA, ScribPSO, and ScribHHO adopt a spreading approach from different starting points like the centre or a corner. This diversity allows users to select an approach that aligns with their artistic vision. The number of iterations reflects personal taste and artistic vision, with some favouring simplicity and abstraction while others seek intricate detail. Thus, choosing an algorithm and iteration count is more than technical. It underscores how individual preferences shape the creative outcome, akin to how different artists have distinctive styles and techniques.

Figure \ref{fig:compared} presents the qualitative results of scribble drawings generated using various backbone algorithms. While the visual outcomes appear similar, each algorithm introduces subtle artistic variations. To highlight these, we plotted the convergence curves (see Figure \ref{fig:quant}, top row) for different test images, showing how each method progressively improves the fitness score. Additionally, Figure \ref{fig:quant} (bottom row) displays the Structural Similarity (SSIM) \cite{sara2019image} between the generated results and the edge maps from the Laplacian of Gaussian (LoG) applied to the input images.  All algorithms captured over 80\% of the visual information, with ScribGA achieving nearly 90\%, making it the top performer. For all subsequent comparisons, we refer to ScribGA as ScribGen. 

\begin{figure*}[!htb]
\centering
\Description{The figure shows the comparative result of the proposed ScribGA with methods proposed by Chan et al. and Cole et al.}
\includegraphics[width=\textwidth]{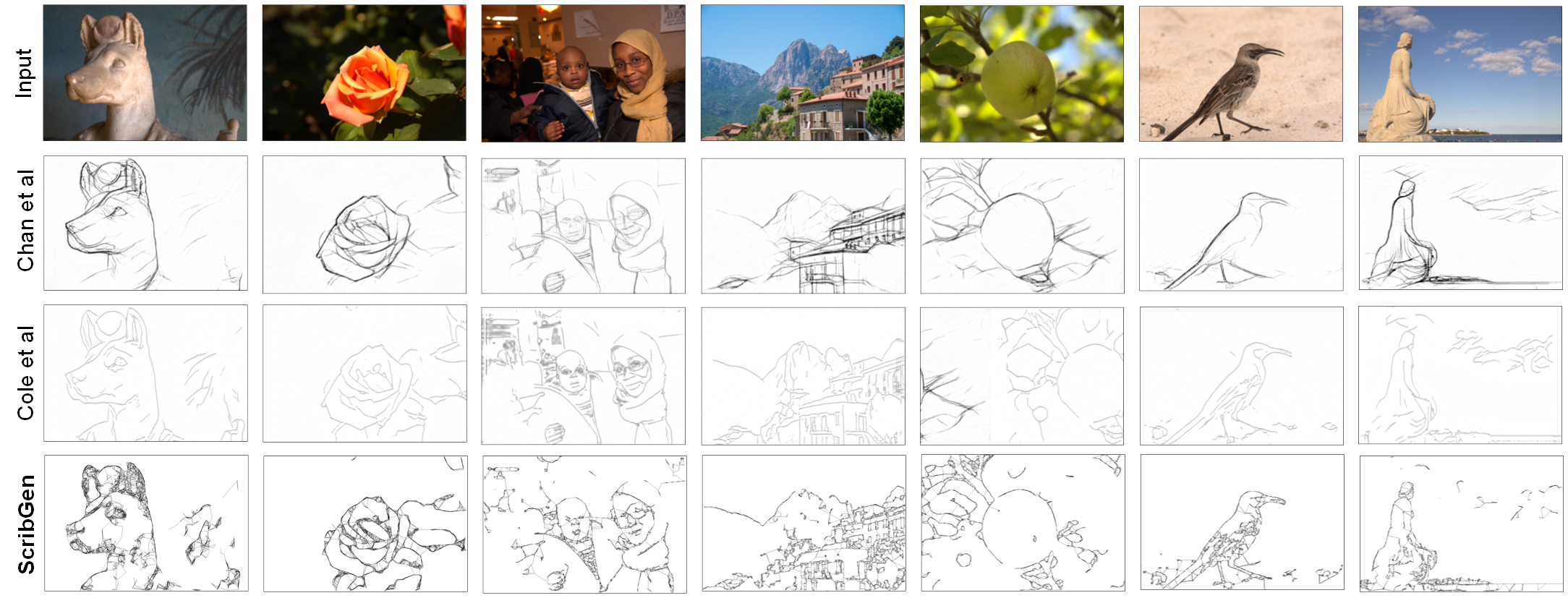}
\caption{Comparison between the ScribGA, \cite{chan2022learning} (Open Sketch mode) and \cite{cole2008people} over images from COCO \cite{lin2014microsoft} and  MIT-Adobe FiveK \cite{fivek} datasets.}
\label{fig:onlyChenCompared}
\end{figure*}

\begin{figure*}[h]
\centering
\Description{The figure compares our results across two styles: open sketch (rows 1 and 2) and contour drawing (rows 3 and 4). Notably, ScribGen effectively approximates contour drawings by simply adjusting line thickness, which presents an interesting avenue for further research in this field.}
\includegraphics[width=\textwidth]{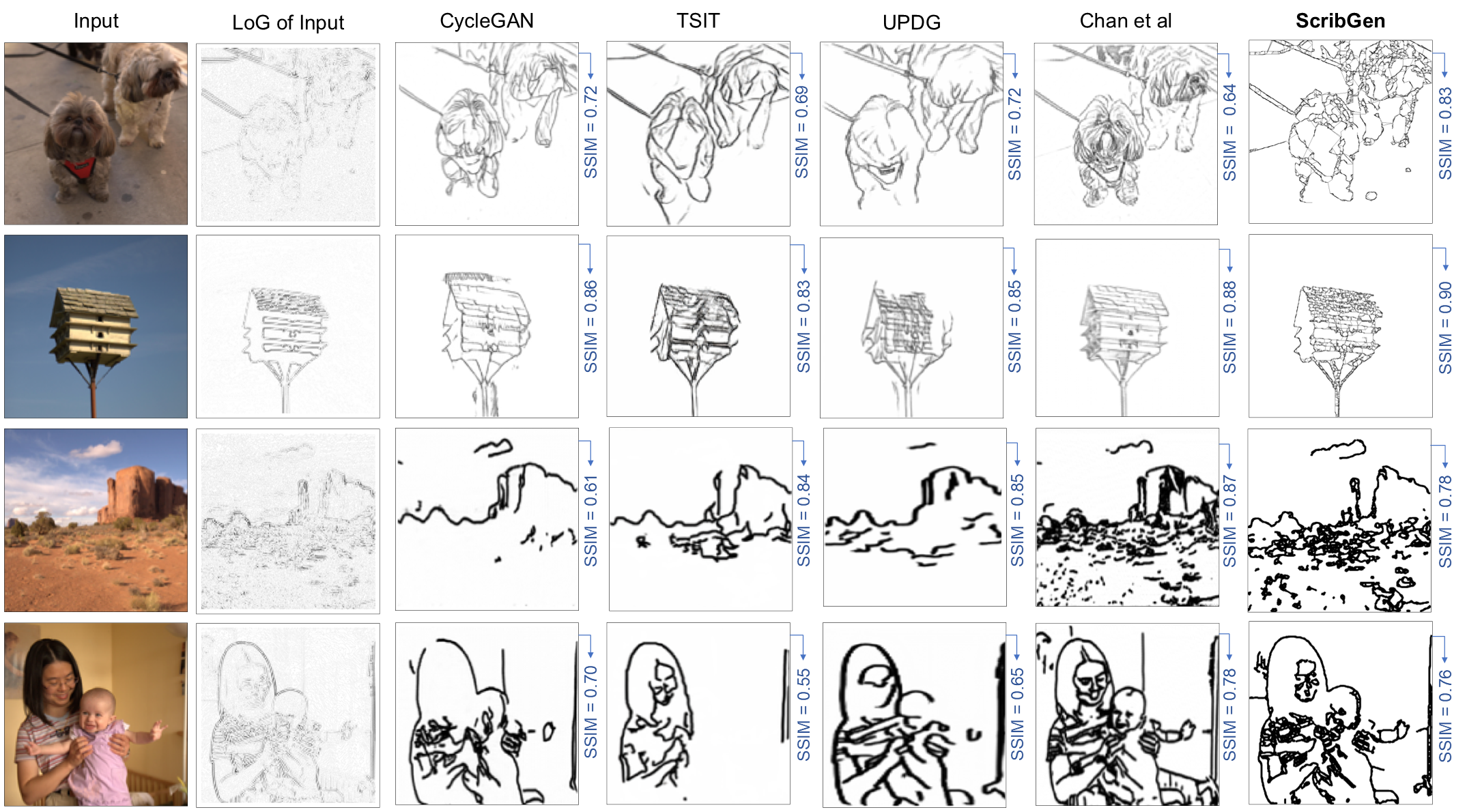}
\caption{Comparison between ScribGA, CycleGAN, TSIT, UPDG, and \cite{chan2022learning} over images from COCO \cite{lin2014microsoft} and  MIT-Adobe FiveK \cite{fivek} datasets.} \label{fig:variousDLCompared}
\end{figure*}

\begin{figure*}[htb]
\centering
\Description{The figure illustrates the diverse applications of ScribGen across various domains.}
\includegraphics[width=0.95\textwidth]{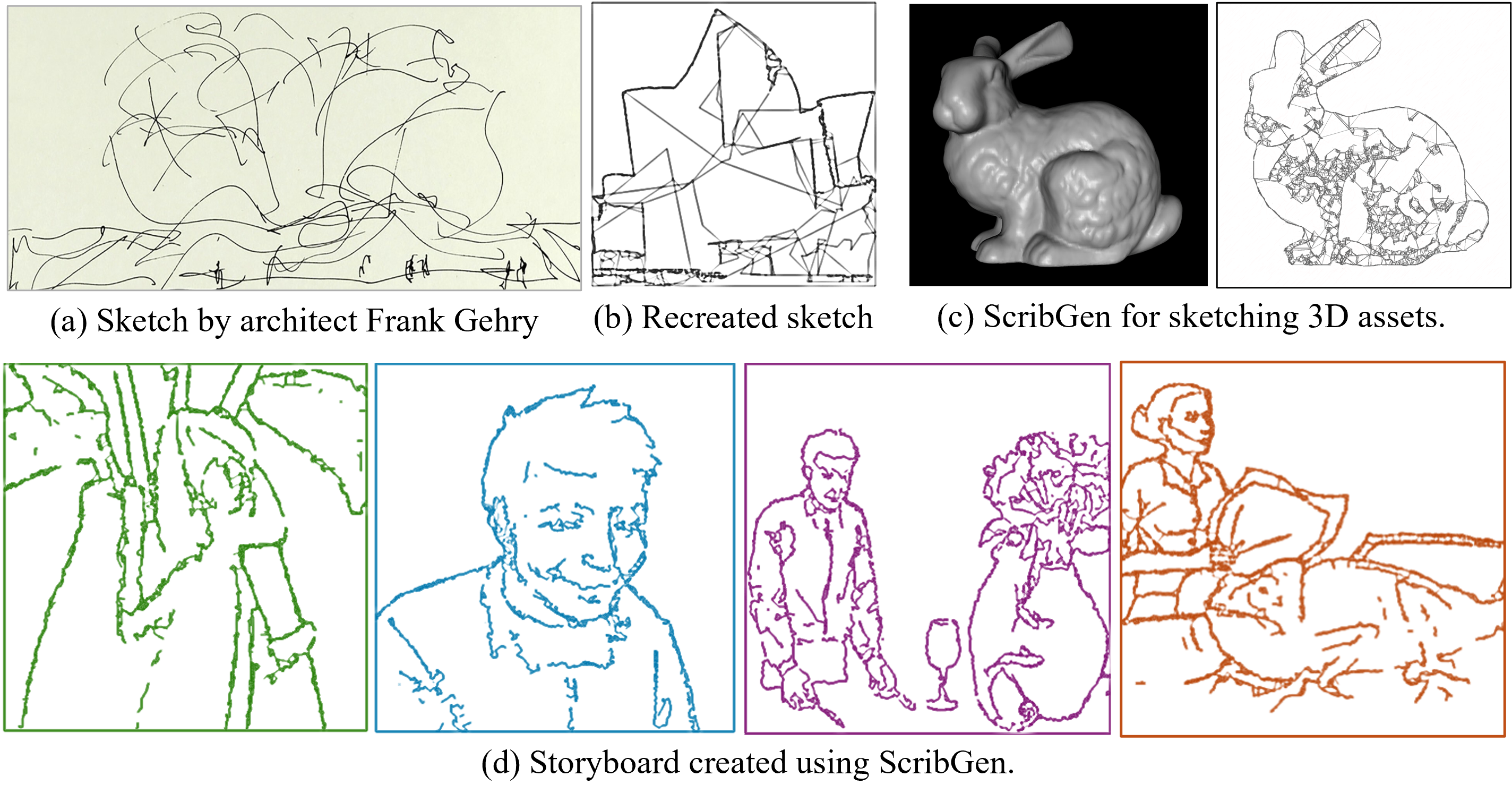}
\caption{Applications of ScribGen. Image credits: (a) \cite{charitonidou2022frank}; (c) Stanford 3D Scanning Repository [graphics.stanford.edu].} 
\label{fig:application}
\end{figure*}

\subsection{Single Continuous vs Composite Generation}

Figure \ref{fig:combination} compares scribbles generated by ScribGen over 2000 generations with a composite image from four separate runs of 500 generations each. We can observe better visual appeal of the composite image.

\textbf{Why does compositing yield better results?} Composition accounts for the benefits of learning through different evolutionary paths. In continuous settings, the population may gravitate towards a specialized skill or feature, emphasizing certain traits. Conversely, blending the outputs of four separate evolutionary runs combines different characteristics. Multiple evolutionary paths introduce a broader range of visual features, leading to more compelling and realistic outcomes.

Users can choose between the two scribbling strategies, i.e. continuous or composite. Continuous evolution offers gradual refinement, appealing to those who value consistency and detailed development. Composite generation combines multiple paths, appealing to users who prefer diversity and complexity. The choice reflects each user's artistic preferences and visual style.

\subsection{Comparison with Deep Generative Networks}
In the era of deep generative AI, the study of any generative algorithm remains incomplete without its comparison with deep generative methods.
\cite{chan2022learning} proposed a GAN-based line drawing method for generating line sketches under different styles such as contour drawing, and open sketch. However, their method relies on supervised training over a dataset of images and a dataset of differently styled line drawings. 

Figure \ref{fig:onlyChenCompared} compares the results obtained by ScribGen with the line drawings obtained by \cite{chan2022learning} and  \cite{cole2008people} over widely different images. We observe that ScribGen's performance is better than that of \cite{cole2008people} and is nearly comparable to that of \cite{chan2022learning}. It is important to observe that metaheuristic-based methods require no ground truth supervised training simply due to their optimization-based setup, and they scale very well to various scenes.

Figure \ref{fig:variousDLCompared} compares our approach to four unpaired image-to-image translation methods for the image-to-line drawing task such as CycleGAN \cite{zhu2017unpaired}, TSIT \cite{jiang2020tsit}, Unpaired Portrait Drawing Generation (UPDG) \cite{yi2020unpaired}, and \cite{chan2022learning}. 

We compare our results under two different styles: open sketch (Figure \ref{fig:variousDLCompared}, row: 1,2) and contour drawing (Figure \ref{fig:variousDLCompared}, row: 3,4). It is interesting to note that 
just by adjusting line thickness, ScribGen effectively approximates contour drawings, presenting a potential future research direction. We report SSIM to compare the similarity between results and the edge maps from Laplacian of Gaussian (LoG) over input images, with ScribGen achieving the highest SSIM. This shows that generating "meaningfully random" drawings from colour images is closely tied to heuristic-based rather than learning-based methods due to the lack of training data in this domain. 

\section{Applications}
Scribble art has applications in therapy \cite{mcnamee2004using}, stress relief, education, design, animation, arts, marketing, and architectural renderings. Figure \ref{fig:application} (a) showcases architect Frank Gehry's initial concept sketch for the Walt Disney Concert Hall \cite{charitonidou2022frank}. Figure \ref{fig:application} (b) illustrates scribble art generated by ScribGen for the hall, aligning closely with the completed architecture [\url{https://www.laphil.com/about/our-venues/about-the-walt-disney-concert-hall}]. A comparison reveals that the artist’s scribble art shows more irregular lines, while ScribGen produces more structured designs. ScribGen can be used for sketching 3D objects, shown in Figure \ref{fig:application} (c) \cite{turk1994zippered}. It is noteworthy how ScribGen captures the features of the input image, including the nuanced lighting effects. Additionally, ScribGen extends to the creation of artistic text-works, as shown in Figure \ref{fig:teaser}, and can be utilized in redesigning user interfaces and logos, depicted in Figure  \ref{fig:combination} (Column 1).  ScribGen also integrates into concept art and story-boarding as shown in Figure \ref{fig:application} (d). 

\section{Conclusion}
The work explored metaheuristic optimization for scribble art generation, showing its versatility and simplicity compared to deep generative methods. Our approach requires no supervised training or ground truth datasets and often surpasses established deep learning techniques in maintaining visual similarity and detail. This demonstrates the potential of metaheuristic algorithms as a robust solution for generating scribble drawings from colour images. Future research aims to enhance the efficiency of metaheuristic algorithms in parameter optimization for art generation, extending its applicability to styles such as line art and poly art. Additionally, we plan to use this framework as a data source for training and fine-tuning deep networks in sketch-based image generation.  

\begin{acks}
We acknowledge the generous support from the Prime Minister Research Fellowship (PMRF) grant and the Jibaben Patel Chair in Artificial Intelligence. We also thank Sakshi, IIT Gandhinagar, for her valuable inputs during this project.
\end{acks}

\bibliographystyle{ACM-Reference-Format}
\bibliography{bibfile}

\end{document}